\documentclass[pra,twocolumn,superscriptaddress]{revtex4}

\usepackage{graphicx,epic,eepic,epsfig,amsmath,latexsym,amssymb,verbatim,subfigure,color}
\usepackage{theorem}

\newtheorem{definition}{Definition}

\newtheorem{lemma}[definition]{Lemma}

\newtheorem{theorem}[definition]{Theorem}

\def\squareforqed{\hbox{\rlap{$\sqcap$}$\sqcup$}}
\def\qed{\ifmmode\squareforqed\else{\unskip\nobreak\hfil
\penalty50\hskip1em\null\nobreak\hfil\squareforqed
\parfillskip=0pt\finalhyphendemerits=0\endgraf}\fi}
\def\endenv{\ifmmode\;\else{\unskip\nobreak\hfil
\penalty50\hskip1em\null\nobreak\hfil\;
\parfillskip=0pt\finalhyphendemerits=0\endgraf}\fi}
\newenvironment{proof}{\noindent \textbf{{Proof~} }}{\qed}

\mathchardef\ordinarycolon\mathcode`\:
\mathcode`\:=\string"8000
\def\vcentcolon{\mathrel{\mathop\ordinarycolon}}
\begingroup \catcode`\:=\active
  \lowercase{\endgroup
  \let :\vcentcolon
  }

\newcommand{\nc}{\newcommand}
\nc{\rnc}{\renewcommand}
\nc{\beq}{\begin{equation}}
\nc{\eeq}{{\end{equation}}}
\nc{\beqa}{\begin{eqnarray}}
\nc{\eeqa}{\end{eqnarray}}
\nc{\lbar}[1]{\overline{#1}}
\nc{\bra}[1]{\langle#1|}
\nc{\ket}[1]{|#1\rangle}
\nc{\ketbra}[2]{|#1\rangle\!\langle#2|}
\nc{\braket}[2]{\langle#1|#2\rangle}
\nc{\proj}[1]{| #1\rangle\!\langle #1 |}
\nc{\avg}[1]{\langle#1\rangle}
\nc{\Rank}{\operatorname{Rank}}
\nc{\smfrac}[2]{\mbox{$\frac{#1}{#2}$}}
\nc{\Tr}{\operatorname{Tr}}
\nc{\tr}{\operatorname{Tr}}
\nc{\id}{\operatorname{id}}
\nc{\1}{\openone}
\nc{\ox}{\otimes}
\nc{\dg}{\dagger}
\nc{\dn}{\downarrow}
\nc{\cA}{{\cal A}}
\nc{\cB}{{\cal B}}
\nc{\cC}{{\cal C}}
\nc{\cD}{{\cal D}}
\nc{\cE}{{\cal E}}
\nc{\cF}{{\cal F}}
\nc{\cG}{{\cal G}}
\nc{\cH}{{\cal H}}
\nc{\cI}{{\cal I}}
\nc{\cJ}{{\cal J}}
\nc{\cK}{{\cal K}}
\nc{\cL}{{\cal L}}
\nc{\cM}{{\cal M}}
\nc{\cN}{{\cal N}}
\nc{\cO}{{\cal O}}
\nc{\cP}{{\cal P}}
\nc{\cR}{{\cal R}}
\nc{\cS}{{\cal S}}
\nc{\cT}{{\cal T}}
\nc{\cX}{{\cal X}}
\nc{\cY}{{\cal Y}}
\nc{\cZ}{{\cal Z}}
\nc{\supp}{{\operatorname{supp}}}
\nc{\var}{\operatorname{var}}
\nc{\rar}{\rightarrow}
\nc{\lrar}{\longrightarrow}
\nc{\polylog}{\operatorname{polylog}}

\def\a{\alpha}

\def\g{\gamma}
\def\d{\delta}
\def\e{\epsilon}

\def\ph{\varphi}

\nc{\RR}{{{\mathbb R}}}
\nc{\CC}{{{\mathbb C}}}
\nc{\FF}{{{\mathbb F}}}
\nc{\NN}{{{\mathbb N}}}
\nc{\ZZ}{{{\mathbb Z}}}
\nc{\PP}{{{\mathbb P}}}
\nc{\QQ}{{{\mathbb Q}}}
\nc{\UU}{{{\mathbb U}}}
\nc{\EE}{{{\mathbb E}}}
\nc{\Icoh}{{I^{\rm coh}}}
\nc{\Qca}{{Q_{\rm ss}}}
\nc{\Qcaa}{{Q^{(1)}_{\rm ss}}}
\nc{\Dcaa}{{D^{(1)}_{{\rm ss}\rightarrow}}}
\nc{\Dca}{{D_{{\rm ss}\rightarrow}}}

\nc{\be}{\begin{equation}}
\nc{\ee}{{\end{equation}}}
\nc{\bea}{\begin{eqnarray}}
\nc{\eea}{\end{eqnarray}}
\nc{\<}{\langle}
\rnc{\>}{\rangle}
\nc{\Hom}[2]{\mbox{Hom}(\CC^{#1},\CC^{#2})}
\nc{\rU}{\mbox{U}}

\begin{document}
\author{Graeme Smith}
\email{gsbsmith@gmail.com}
\affiliation{Department of Computer Science, University of Bristol, Bristol, BS8 1UB, UK}

\title{The private classical capacity with a symmetric  side channel and its application to quantum cryptography}
 %Can a public channel increase the private capacity?}
\date{May 25, 2007}
\parskip=1ex
\parindent=0ex

\begin{abstract}
We study the symmetric-side-channel-assisted private capacity of a quantum channel, 
for which we provide a single-letter formula.  This capacity is additive, convex, and, for degradable channels, equal to the unassisted private capacity.  While a channel's (unassisted) capacity for for private classical communication may be strictly larger than its quantum capacity, 
we will show that these capacities are equal for degradable channels, thus demonstrating the equivalence of privacy and quantum coherence in this context.  We use these ideas to find new bounds on the key rate of quantum key distribution protocols with one-way classical post-processing. For the Bennett-Brassard-84 (BB84) protocol, our results demonstrate that collective attacks are strictly stronger than individual attacks.
\end{abstract}

\maketitle

\section{Introduction}

One of the earliest results in quantum information theory was the realization in \cite{BB84} that a noisy quantum channel can be used to establish secret correlations whose security is guaranteed by the fundamental laws of physics.  Furthermore, while the full-scale implementation of quantum  computation is likely to remain a distant hope for years to come, secure quantum key distribution protocols may begin to play an important role in the world of information security in the not-too-distant future.

In the simplest case of independent and identically distributed noise, which we will consider here and to which most quantum key distribution (QKD) protocols can be reduced \cite{RenThesis, GL03}, the capacity of a channel for private classical communication was studied in \cite{D03}.  In that work, a {\em multi-letter formula} for the private classical capacity was provided (and, indeed, a similar formula for a channel's capacity for quantum communication).  Unfortunately, this multi-letter formula cannot be evaluated in general, and thus provides only a partial characterization of the capacity we seek.

In lieu of a closed form expression for the private classical capacity of a quantum channel, which we will call $C_p$, it is the primary purpose of this work to provide upper bounds.  In particular, we will consider the capacity of a quantum channel for private classical communication when assisted by the family of 
(one-way) quantum channels that map symmetrically to their output and environment.  Since any such channel has zero private capacity on its own, one would expect the resulting bound to be quite tight.  This approach is very much in the spirit of \cite{SSW06}, and in fact our expression for the symmetric side-channel assisted private capacity (ss-private capacity) shares many of the nice properties
 of the ss-capacity found there, namely it is single-letter, additive, and convex.  

A secondary goal of this work is to explore the connection between unconditional privacy and quantum coherence, which has long been folklore in the
quantum information community and provided motivation for the coding
 theorems proved in \cite{D03} (see also \cite{SW98,DW04}).  While this analogy is quite useful, it is 
 known that the correspondence is not exact.
  Indeed, there are quantum channels for which the capacity for private communicaton and quantum communication are quite different.  In \cite{HHHO03} it was shown
 that there exist quantum states from which no entanglement can be distilled via two-way classical 
 communication but which nevertheless 
 can be used to create secure key via one-way public classical communication.   This leads to examples of channels with zero quantum capacity but nonzero private classical capacity.
 
  Understanding the connection between 
  coherence and privacy  in a quantitative way does not seem to be possible at the moment, as there do not yet exist simple expressions for either the private classical or quantum  capacities of a channel.  
 However, we will show below that, for the class of channels known as {\em degradable}, it is possible to find a simple expression for the private classical capacity,  $C_p$, and indeed for such channels $C_p$ is exactly the quantum capacity (which, due to  \cite{DS03} has a closed-form expression).  As well as giving the first examples of nontrivial channels for which $C_p$ can be found explicitly, this provides a
  setting in which the above mentioned analogy between privacy and coherence can be made exact.

 Furthermore, the ss-private-capacity of a degradable channel is  exactly equal to its (unassisted) quantum capacity. We will combine this result with the convexity of the ss-private capacity to provide 
 a new technique for upper-bounding the private capacity of a general quantum channel, extending the current best known bounds for the quantum capacity of the depolarizing to its private capacity and providing new bounds for private capacity of a channel with independent phase and amplitude noise.  This last result leads to collective attacks on  BB84 that outperform the optimal individual attack.

The rest of the paper is organized as follows.  In Section II we study the private capacity 
of a degradable channel, in Section III we provide a single-letter formula for the ss-private
 capacity of a general channel, while in Section IV we provide upper bounds for the private 
 capacity of some specific quantum channels and discuss their relation to collective attacks 
 in QKD.  In Section V we mention a few open problems.

\section{Noisy processing is no help for degradable $\cN$}

In a classical setting, if we imagine a broadcast channel which maps $\cN: X \rightarrow (Y,Z)$, where $Y$ is the output to the receiver and $Z$ is the output of an eavesdropper, it was 
shown in \cite{CK78} that the secret-key capacity of $\cN$ is exactly
\begin{equation}\label{Eq:CK}
C_p(\cN) = \sup_{X\rightarrow T}\left( I(T;Y) - I(T;Z)\right).
\end{equation}
Here the optimization is over a reference variable $X$, which represents the distribution of messages sent through the channel, together with a noisy processing of $X$ that generates $T$.

By analogy with this result, one may imagine that the private classical capacity of a quantum 
channel would be given by
\begin{equation}\nonumber
C^{(1)}_p(\cN) := \sup_{\{p_x, \ket{\ph_x}\}, X\rightarrow T} \left(I(T;B)_\omega - I(T;E)_\omega \right), 
\end{equation}
where  $\omega_{ABE} = \sum_{x,t}p(t|x)p(x)\proj{t}_A \ox U_{\cN}\proj{\ph_x}U_{\cN}^\dg$ with 
$U_\cN$ an isometric extension of $\cN$ (i.e., $\cN(\rho) = \Tr_E U_\cN \rho U_\cN^\dg$).
So, the optimization would again be taken over input random variable $X$ (this time with a choice
 of basis), together with a classical noisy processing $X\rightarrow T$.  Indeed, the coding theorem proved  in \cite{D03} showed that this rate is in fact achievable:

\begin{equation}\nonumber
C_p(\cN) \geq C^{(1)}_p(\cN),
\end{equation}

but did not establish the converse statement.  Instead, it was shown that
\begin{equation}\nonumber
C_p(\cN) = \lim_{n \rightarrow \infty} \frac{1}{n} C^{(1)}_p(\cN^{\ox n}).
\end{equation}
Evidence was found in \cite{SRS06} that this regularization, as the limit over $n$ is typically 
called in this context,  is necessary in general.

A class of channels for which we will be able to explicitly evaluate $C_p$ are called degradable,
 and were defined in \cite{DS03} in analogy with the classical notion of a degraded broadcast 
 channel \cite{Cover72}.

\begin{definition}
A channel $\cN$ is called {\em degradable} if there exists a  completely positive trace preserving 
{\em degrading map} $\cD$ such that
\begin{equation}\nonumber
\cD\circ \cN = \widehat{\cN},
\end{equation}
where ${\cN}(\rho) = \Tr_E U_\cN \rho U_\cN^\dg$ and
 $\widehat{\cN}(\rho) = \Tr_BU_\cN \rho U_\cN^\dg$.
\end{definition}

Below, we will prove  that the private classical capacity of a
degradable channel is equal to its quantum capacity.  The quantum capacity is, in turn, equal to the
single-letter optimized coherent information.  This result is very much in line with the findings of 
\cite{CK78}, in which it was shown that in the classical case if $Z$ is a degraded version of $Y$, the noisy processing in Eq.~(\ref{Eq:CK}) is unecessary.

\begin{theorem}\label{Theorem:DegradableCapacity}
If $\cN$ is degradable, then 
\begin{equation}\nonumber
C_p(\cN) = Q^{(1)}(\cN) =  \sup_{\phi}I(A\rangle B)_{I\ox \cN \proj{\phi}},
\end{equation}
where $I(A\rangle B)_{\rho} = S(B) - S(AB)$.
\end{theorem}

To prove this, we will need the following lemma.
\begin{lemma}\label{Lemma:KoneIsQone}
If $\cN$ is degradable, then
\begin{equation}\nonumber
C^{(1)}_p(\cN) = Q^{(1)}(\cN) = \sup_{\phi}I(A\rangle B)_{I\ox \cN \proj{\phi}}.
\end{equation}
\end{lemma}
\begin{proof}
Let $\cN$ be degradable, fix $\phi = \sum_x p_x \proj{\ph^x}$ and let
\begin{equation}\nonumber
\omega_{XTBE} = \sum_{x,t}p_{x,t} \proj{x}_X\ox \proj{t}_T \ox U_{\cN}\proj{\ph_x}U_\cN^\dg.
\end{equation}
Then  $I(X;B) = I(T;B) + I(X;B | T)$,
which is a consequence of the chain rule, together with the fact that $I(XT;B) = I(X;B)$ because $X\rightarrow T$.  This implies that
\begin{eqnarray}\nonumber
C^{(1)}_p(\cN) &=& \sup_{\{p_x, \ket{\ph_x}\}, X\rightarrow T} \left(I(T;B)_\omega - I(T;E)_\omega \right)\\
 & = & \sup_{\{p_x, \ket{\ph_x}\}, X\rightarrow T} \Bigl( I(X;B) - I(X;E) \nonumber \\
& & - \left( I(X;B | T) - I(X; E | T)\right) \Bigr).\nonumber
\end{eqnarray}
Since $\cN$ is degradable, and conditional mutual information is monotonic under local operations (LO) 
when the system conditioned on is classical (an immediate consequence of the LO monotonicity of mutual information, itself a consequence of strong subadditivity), we have $I(X;B | T) \geq I(X; E | T)$,
so that
\begin{eqnarray}
C^{(1)}_p(\cN) &=&  \sup_{\{p_x, \ket{\ph_x}\}} \left( I(X;B) - I(X;E) \right) \nonumber \\
& = & \sup_{\{p_x, \ket{\ph_x}\}}\left( S(B) - S(B|X) - S(E) + S(E|X)\right)\nonumber \\
& = & \sup_{\phi}\left(S(B)-S(E)\right) \nonumber \\
& = & \sup_{\phi}I(A\rangle B) = Q^{(1)}(\cN).\nonumber
\end{eqnarray}
\end{proof}

The following lemma, which shows that $Q^{(1)}$ is additive for degradable channels, was proved in \cite{DS03}.  We provide an alternate proof for both clarity and completeness.

\begin{lemma}\label{Lemma:Q1AdditiveForDegradable}
For $\cN_1$ and $\cN_2$  degradable,
\begin{equation}\nonumber
Q^{(1)}(\cN_1 \ox \cN_2) = Q^{(1)}(\cN_1) + Q^{(1)}(\cN_2).
\end{equation}
\end{lemma}
\begin{proof}
Let $\ket{\phi}_{AA^\prime_1 A^\prime_2}$ be optimal for $Q^{(1)}(\cN_1 \ox \cN_2)$, namely
\begin{equation}\nonumber
Q^{(1)}(\cN_1 \ox \cN_2)  = I(A\rangle B_1B_2)_{I \ox \cN_1\ox \cN_2 \proj{\phi}}.
\end{equation} 
We would like to show that
\begin{equation}\label{Eq:DegradCoherentInfAdd}
I(AA^\prime_1 \rangle B_2) + I(AA^\prime_2 \rangle B_1)  \geq I(A \rangle B_1B_2),
\end{equation}
since this would immediately imply $Q^{(1)}(\cN_1) + Q^{(1)}(\cN_2) \geq Q^{(1)}(\cN_1\ox \cN_2)$, 
and therefore the theorem.

In fact, Eq.~(\ref{Eq:DegradCoherentInfAdd}) is equivalent to 
\begin{equation}\nonumber
I(B_1;B_2) \geq I(E_1;E_2),
\end{equation}
which, is satisfied due to the degradability of $\cN_1$ and $\cN_2$ together with the monotonicity of 
mutual information under local operations.
\end{proof}

We are now in a position to prove Theorem 2.

\begin{proof}{\bf [of Theorem 2]}
Let $\cN$ be degradable.  Then, from \cite{D03}, the secret-key capacity of $\cN$ is
\begin{equation}\nonumber
C_p(\cN) = \lim_{n\rightarrow \infty}\frac{1}{n}C^{(1)}_p(\cN^{\ox n}).
\end{equation}

By Lemma \ref{Lemma:KoneIsQone} and the degradability of $\cN^{\ox n}$, we have
\begin{equation}\nonumber
C_p(\cN) = \lim_{n\rightarrow \infty}\frac{1}{n}Q^{(1)}(\cN^{\ox n}),
\end{equation} 
while Lemma \ref{Lemma:Q1AdditiveForDegradable} gives us $C_p(\cN) = Q^{(1)}(\cN)$.
\end{proof}

\section{Private classical capacity with a symmetric side-channel}
Before defining the capacity to be studied, we must first formally define the notion of a private classical
 code.  An $(n,K)$ key code, $C$, is a set of $K$ states on $A^{\ox n}$, together with a 
 decoding operation 
$\cD_n: \cB(B^{\ox n}) \rightarrow  \{1, \dots K\}$.  The rate of such a code is $(\log K)/n$.  Such a code is called $\e$-good for a channel $\cN^{(n)}$ (mapping $A^{\ox n}$ to $B^{\ox n}$)
if, defining 
\begin{equation}\nonumber
\rho_{A B^{\ox n} E^{\ox n}} = \frac{1}{K}\sum_{x=1}^K \proj{x}_{A} \ox U_{\cN^{(n)}}\rho_x (U_{\cN^{(n)}})^\dg,
\end{equation}
we have
\begin{equation}\nonumber
|| I_A \ox \cD_n \ox I_{E^{\ox n}} (\rho_{AB^{\ox n}E^{\ox n}}) - \frac{1}{K} \sum_{x = 1}^K \proj{x}\ox\proj{x} \ox \rho_E ||_1 < \e.
\end{equation}

We say that a rate $R$ is achievable over $\cN^{\ox n}$ if for every $\e > 0$ and all sufficiently large $n$ there is a 
a code $C_n \subset A^{\ox n}$ that is $\e$-good for $\cN^{\ox n}$ 
with $\lim_{n \rightarrow \infty} \frac{\log |C_n|}{n} \geq R$.
The private classical capacity of $\cN$ is then defined as the maximum achievable rate.

Letting $S = S_d \subset \top \ox \bot$ be the $d(d+1)/2$-dimensional symmetric subspace between $\top$ and $\bot$ and $V_d: \CC^{d(d+1)/2} \rightarrow S$, we call
\begin{equation}\nonumber
\cA_d(\rho) = \Tr_{\bot}V_d \rho V_d^\dg
\end{equation}
the {\em d-dimensional symmetric channel}.  Note that $\cA_d$ maps states on $\CC^{d(d+1)/2}$ to states on $\CC^d$.

The symmetric side-channel assisted private classical capacity of a channel $\cN$ is 
simply the private capacity of $\cN$ when assisted by an arbitrary symmetric channel.  
More formally, we say that a rate $R$ is ss-achievable 
if for all $\e>0$ and sufficiently large $n$ there is a $d_n$ such that $R$ is $\e$-achievable over
$\cN^{\ox n}\ox \cA_{d_n}$.  The ss-private classical capacity is then the maximum ss-achievable key rate.  The main result of this work is the following theorem characterizing the ss-private capacity.

\begin{theorem}\label{Thm:SSCAP}
The ss-private capacity of $\cN$ is
\begin{equation}\label{Eq:Char1}
C^{(1)}_{p,ss}(\cN) = \sup_{\{p_x,\ket{\ph_x}_{AFG}\} X\rightarrow T} \left( I(T;BF) - I(T;EG)\right),
\end{equation}
with the optimization over $\ket{\ph_x}_{AFG}$ symmetric in $FG$. 
\end{theorem}

Note that this expression for $C_{p,ss}$ is related to but differs from the upper bound presented in \cite{KGR05}, which in this case translates to
\begin{equation}\nonumber
 \sup_{\{p_x,\ket{\ph_x}_{A} \} X\rightarrow \sigma_U, X\rightarrow\sigma_V} \left( I(U;BV) - I(U;EV)\right)
\end{equation}
 in that the optimization in Theorem \ref{Thm:SSCAP} is restricted to classical $T$ rather than a general $\sigma_U$.  So that besides admitting an operational interpretation, our bound will in general be tighter.

A useful alternative characterization of $C_{p,ss}$ is given by
\begin{equation}\label{Eq:LimitKoneCharacterization}
C^{(1)}_{p,ss}(\cN) = \sup_d C^{(1)}_p(\cN \ox \cA_d),
\end{equation}
which can be seen to be equivalent to Eq.~(\ref{Eq:Char1}) as follows.  To see that Eq.~(\ref{Eq:Char1})
can be no bigger than Eq.~(\ref{Eq:LimitKoneCharacterization}), note that any ensemble 
$\{ p_x, \ket{\ph_x}_{AFG}\}$ with $\ket{\ph_x}_{AFG}$ symmetric in $FG$ can be generated using $\cA_{d}$ with $d = d_F$.  Alternatively, given any ensemble of states, $\{ p_x, \ket{\ph_{AS_d}}\}$, where $S_d$ is the input to $\cA_d$, we retrieve an ensemble $\{ p_x, I\ox U_{\cA_d}\ket{\ph_{AS_d}}\}$ 
which is symmetric in $FG$, so that Eq.~(\ref{Eq:Char1}) is no smaller than Eq.~(\ref{Eq:LimitKoneCharacterization}).

Before proving the theorem, we provide a multi-letter characterization of the capacity.

\begin{lemma}\label{Lemma:MultiLetterKss}
\begin{equation}\nonumber
C_{p,ss}(\cN) = \lim_{n \rightarrow \infty} \frac{1}{n}C^{(1)}_{p,ss}(\cN^{\ox n})
\end{equation}

\end{lemma}
\begin{proof}
To see that the ss-private capacity is no less than the right-hand side, note that for any ensemble $\{ p_x, \ket{\ph_x}_{A^nFG}\}$ symmetric in $FG$ and $X \rightarrow T$, a rate of
\begin{equation}\nonumber
\frac{1}{n}\left( I(T; B^nF) - I(T; E^n G) \right)
\end{equation}
is achievable by the coding theorem of \cite{D03}.  

Conversely, fix $\e > 0$, let $\{\frac{1}{2^{nR}}, \rho^k_{(A^\prime)^n FG} \}$ 
be an $(n,\e)$ ss-private code, 
and let
\begin{equation}\nonumber
\omega = \frac{1}{2^{nR}}\sum_{k = 1}^{2^{nR}}\proj{k}_T \ox \rho^k_{(A^\prime)^n FG}.
\end{equation} 
Then, letting $\cD$ be the decoding operation associated with the code,

\begin{equation}\nonumber
\sigma = \left( I_T \ox U_{\cN}^{\ox n} \ox I_{FG} \right)\omega  \left( I_T \ox (U_{\cN}^\dg)^{\ox n} \ox I_{FG}\right) ,
\end{equation}
and
\begin{equation}\nonumber
\rho = I_T \ox \cD_{B^n} \ox I_{EFG} (\sigma),
\end{equation}

we have
\begin{equation}
\Biggl|\Biggl| \rho -  \frac{1}{2^{nR}} \sum_{k=1}^{2^{nR}} \proj{k}_T \ox \proj{k}_{C} \ox \rho_{EG} \Biggr|\Biggr|_1 \leq \e. \nonumber
\end{equation}

As a result, 

\begin{eqnarray} \nonumber
I(T; B^nF)_{\sigma_{TB^n F}} 
& \geq & I(T; C)_{\rho_{TC}}  \nonumber \\
& \geq & n R - 2\left( 2\e nR + H(\e)\right),\nonumber
\end{eqnarray}
where in the last line we have used the continuity result of \cite{AF03}.
Similarly, since $\sigma_{TE^nG} \approx \sigma_T \ox \sigma_{E^nT}$, we have
\begin{equation}\nonumber
I(T; E^nG)_{\sigma_{TE^n G}} \leq 2\left( 2\e nR + H(\e)\right),
\end{equation}
so that

\begin{eqnarray}\nonumber
I(T; B^nF)_{\sigma_{TB^nF}} - I (T; E^n G)_{\sigma_{TE^nG}} \nonumber\\
 \geq  nR - 4\left( 2\e nR + H(\e)\right) \nonumber\\
 =  n R(1- 8\e) - 4H(\e).\nonumber
\end{eqnarray}
Thus,
\begin{equation}\nonumber
R \leq \frac{1}{1 - 8\e}\left( \frac{1}{n}C^{(1)}_{p,ss}(\cN^{\ox n}) + 4H(\e)\right).
\end{equation}
\end{proof}

Now using the following lemma, which shows that $C^{(1)}_{p,ss}$ is additive, we will be in a position to 
prove Theorem \ref{Thm:SSCAP}.

\begin{lemma}\label{Lemma:AdditiveKss}
$C_{p,ss}^{(1)}$ is additive:
\begin{equation}\nonumber
C^{(1)}_{p,ss}\left(\cN_1 \ox \cN_2\right) = C^{(1)}_{p,ss}\left(\cN_1\right) + C^{(1)}_{p,ss}\left(\cN_2\right). 
\end{equation}
\end{lemma}
\begin{proof}
For any $\ket{\phi_x}_{A_1A_2FG}$ symmetric in $FG$ and $X\rightarrow T$, let 

\begin{equation}\nonumber
\ket{\phi_x^1}_{A_1B_2E_2FGC_1C_2} =
\end{equation}
\begin{equation}
\frac{1}{\sqrt{2}}\Bigl(I_{A_1}\ox U_{\cN_2}\ox I_{FG}\ket{\phi_x}\ket{01}_{C_1C_2}  + \nonumber
\end{equation}
\begin{equation}
 \left(I_{A_1}\ox{\rm SWAP}_{B_2E_2}\ox I_{FG}\right)I_{A_1}\ox U_{\cN_2}\ox I_{FG}\ket{\phi_x}\ket{10}_{C_1C_2} \Bigr),\nonumber,
\end{equation}
and $\phi^1 = \sum_{x,t}p(x,t) \proj{t}\ox \phi^1_x$.

Then, labeling $\tilde{F}_1 = B_2FC_1$ and $\tilde{G}_1 = E_2GC_2$, we have
\begin{eqnarray}
C_{p,ss}^{(1)}(\cN_1) \geq I(T; B_1\tilde{F}_1)_{I\ox \cN_1 \ox I (\phi^1)}-I(T;E_1\tilde{G}_1)_{I \ox \cN_1 \ox I (\phi^1)}\nonumber\\
= \frac{1}{2}\Biggl( I(T;B_1B_2 F)_{I\ox \cN_1\ox\cN_2 \ox I (\phi)} + I(T;B_1E_2 F)_{I\ox \cN_1\ox\cN_2 \ox I (\phi)} \nonumber\\
- I(T;E_1B_2G)_{I\ox \cN_1\ox\cN_2 \ox I (\phi)} - I(T;E_1E_2 G)_{I\ox \cN_1\ox\cN_2 \ox I (\phi)} \Biggr)\nonumber.
\end{eqnarray}
Similarly defining $\ket{\phi^2_x}$, we find
\begin{eqnarray}
C_{p,ss}^{(1)}(\cN_2) \geq \nonumber I(T; B_2\tilde{F}_2)_{I\ox \cN_2 \ox I (\phi^2)}-I(T;E_2\tilde{G}_2)_{I \ox \cN_2 \ox I (\phi^2)}\\ 
=  \frac{1}{2}\Biggl( I(T;B_1B_2 F)_{I\ox \cN_1\ox\cN_2 \ox I (\phi)} + I(T;E_1B_2 F)_{I\ox \cN_1\ox\cN_2 \ox I (\phi)} \nonumber\\
- I(T;B_1E_2G)_{I\ox \cN_1\ox\cN_2 \ox I (\phi)} - I(T;E_1E_2 G)_{I\ox \cN_1\ox\cN_2 \ox I (\phi)} \Biggr),\nonumber
\end{eqnarray}
so that
\begin{eqnarray}
C_{p,ss}^{(1)}(\cN_1) + C_{p,ss}^{(1)}(\cN_2) \geq
 \ \ \ \ \ \ \ \ \ \ \ \ \ \ \ \ \ \ \ \ \ \ \ \ \ \ \ \ \ \ \ \ \ \ \ \ \ \ \  \nonumber\\
I(T;B_1B_2 F)_{I\ox \cN_1\ox\cN_2 \ox I (\phi)} -  I(T;E_1E_2 G)_{I\ox \cN_1\ox\cN_2 \ox I (\phi)} .\nonumber
\end{eqnarray}
Since this is true for any $\ket{\phi_x}$, we have
\begin{equation}
C_{p,ss}^{(1)}(\cN_1) + C_{p,ss}^{(1)}(\cN_2) \geq C_{p,ss}^{(1)}(\cN_1 \ox \cN_2). \nonumber
\end{equation}
\end{proof}

\begin{proof}{\bf [of Theorem \ref{Thm:SSCAP}]}
By Lemma \ref{Lemma:MultiLetterKss}, we have
\begin{equation}\nonumber
C_{p,ss}(\cN) = \lim_{n \rightarrow \infty}\frac{1}{n}C^{(1)}_{p,ss}(\cN^{\ox n}),
\end{equation}
whereas Lemma \ref{Lemma:AdditiveKss} implies $C^{(1)}_{p,ss}(\cN^{\ox n}) = n C^{(1)}_{p,ss}(\cN)$, 
which gives the result.
\end{proof}

We now show that $C_{p,ss}$ is convex, a property that the unassisted private classical 
capacity is not known to possess. 

\begin{lemma}
$C_{p,ss}$ is convex:
\begin{equation}\nonumber
C_{p,ss}((1{-}p)\cN_0{+}p\cN_1) \leq (1{-}p)C_{p,ss}(\cN_0){+}pC_{p,ss}(\cN_1).
\end{equation}
\end{lemma}

\begin{proof}
Letting $\cN = (1-p)\cN_0\ox\proj{0}_{B_2} + p\cN_1\ox \proj{1}_{B_2}$, we consider the purification of $\cN$ that gives Eve the which channel information in a system $E_2$.   Noting that for any $\cN$ and 
$\cM$, $C_{p,ss}(\cN) \geq C_{p,ss}(\cM\circ\cN)$, we have
\begin{equation}
C_{p,ss}\left( (1-p)\cN_0+p\cN_1\right)\nonumber\\
\end{equation}
\begin{eqnarray}
& \leq & C_{p,ss}\left( (1-p)\cN_0\ox \proj{0}_{B_2}+p\cN_1\ox\proj{1}_{B_2}\right) \nonumber\\
 &= & \sup_{\{p_x,\ket{\ph^x}\} , X\rightarrow T}\left( I(T; BB_2F) - I(T;EE_2G)\right)\nonumber\\
 &=&  \sup_{\{p_x,\ket{\ph^x}\}, X\rightarrow T} \left(\sum_{\a=0}^1 p_\a \left[S(T|EG, \a) - S(T|BF, \a)\right]\right)\nonumber\\
&\leq&   \sum_{\a=0}^1p_\a \sup_{\{p_x,\ket{\ph^x}\}, X\rightarrow T} \left(S(T|EG, \a) - S(T|BF, \a)\right)\nonumber\\
& = & \sum_{\a=0}^1 p_\a C_{p,ss}(\cN_\a),\nonumber
\end{eqnarray}
where the $\ket{\ph^x}$ are taken to be on $A^\prime FG$ and symmetric in $FG$ throughout.
\end{proof}

Finally, we demonstrate that the ss-private-capacity of a degradable channel is, in fact, the single-letter optimized coherent information.

\begin{lemma}
If $\cN$ is degradable, $C_{p,ss}(\cN) = Q^{(1)}(\cN)$.
\end{lemma}
\begin{proof}
For any $d$ and degradable $\cN$, it is also the case that $\cN \ox \cA_d$ is degradable.  As a result, 
\begin{eqnarray}
C^{(1)}_p(\cN\ox \cA_d) &=& Q^{(1)}(\cN\ox\cA_d) \nonumber \\
&=& Q^{(1)}(\cN) + Q^{(1)}(\cA_d) = Q^{(1)}(\cN),\nonumber
\end{eqnarray}
so that, by the characterization of $C_{p,ss}$ in Eq.(\ref{Eq:LimitKoneCharacterization}), we 
have $C_{p,ss}(\cN) = Q^{(1)}(\cN)$.
\end{proof}

We will now use the convexity of $C_{p,ss}$ to show that the following quantity, which we call the 
{\em cost of degradable mixing}, is an upper bound for $C_p$.

\begin{definition}
We define the {\em cost of degradable mixing} as
\begin{equation}\nonumber
C_{\rm DM}(\cN) = \inf_{\{p_i,\cN_i,\cD_i\}} \sum_{i}p_i Q^{(1)}(\cN_i), 
\end{equation}
where the infimum is over $\{p_i,\cN_i,\cD_i\}$ such that
\begin{equation}\nonumber
\cN = \sum_i p_i \cD_i\circ\cN_i
\end{equation}
and each $\cN_i$ is either degradable or anti-degradable.
\end{definition}

That is, we will prove the following theorem.

\begin{theorem}\label{Thm:DMUB}
The cost of degradable mixing of a quantum channel is an upper bound for its private classical capacity.  In other words, $C_p(\cN) \leq C_{\rm DM}(\cN)$.
\end{theorem}

Notice that, by restricting our $\cN_i$ to be either the identity channel or be both degradable and 
anti-degradable, we would recover the upper bound of \cite{MCL06}.

\begin{proof}
Let $\cN = \sum_i p_i \cD_i \circ \cN_i$ be a decomposition of $\cN$ with each $\cN_i$ either degradable or anti-degradable.
Then, noting that $C_p(\cN) \leq C_{p,ss}(\cN)$, and using the convexity of $C_{p,ss}$, we have
\begin{eqnarray}\nonumber
C_p(\cN) & \leq & C_{p,ss}\left( \sum_i p_i \cD_i \circ \cN_i\right) \nonumber\\
 & \leq & \sum_i p_i C_{p,ss}\left(\cD_i \circ \cN_i\right)\nonumber\\
 & \leq & \sum_i p_i C_{p,ss}\left(\cN_i\right)\nonumber\\
 & = & \sum_i p_i Q^{(1)}(\cN_i),\nonumber
\end{eqnarray}
where in the last line we have used the fact that for $\cN_i$ either degradable or antidegradable, $C_{p,ss}(\cN_i) = Q^{(1)}(\cN_i)$.
\end{proof}

One might wonder about the inclusion of $\cD_i$s in the definition of the cost of degradable mixing---wouldn't the bound be tighter if they were all chosen to be the identity?  The trouble is that not 
all channels can be written as a convex combination of degradable and anti-degradable channels, but {\em any} channel can be decomposed into the form required by our definition (e.g., choose only one term, and let $\cN_1 = I$ and $\cD_1 = \cN$, the channel of interest ). 
 In particular, while all extremal qubit channels are either degradable or
  antidegradable (or both)\cite{WPG06}, and therefore any qubit channel can be written as a convex combination
   of such channels, the same is not true in higher dimension.  For example\footnote{Thanks to Debbie Leung for providing this example.}, the tensor product of two extremal qubit channels, 
 one degradable and the other anti-degradable (but neither both), is generically an extremal 
 channel on two qubits, but is neither degradable nor anti-degradable, and in light of its
  extremality cannot be decomposed into such channels. To get around this, we include
  the $\cD_i$s in the definition of the cost of degradable mixing. In the two qubit example,
   we find $C_{DM}$ is exactly equal to the quantum capacity 
of the degradable channel, and therefore so is the private capacity, incidentally  
providing an example of a nondegradable channel for which the private and quantum capacities coincide.

\section{Some specific channels}

Theorem \ref{Thm:DMUB} gives us a technique for bounding the private capacity of a general
 channel, $\cN$, in 
terms of the private capacity of a set of degradable channels appearing in a convex decomposition of 
$\cN$.  We now use this method to provide upper bounds for the key capacity of two channels 
of interest for quantum key distribution --- the Pauli channel with independent phase and 
amplitude noise and the depolarizing channel.  The resulting bounds meet 
or exceed all previously known bounds on the private classical capacity of these channels 
\cite{FGGNP97,MCL06,Bruss98,BG99}.

\subsection{Degradable Channels}
In this subsection we explicitly evaluate the private capacity of some degradable channels.

A qubit channels with two Kraus operators has, up to local unitaries, Kraus operators equal to \cite{RSW02}
\begin{equation}\nonumber
A_0 = \left( \begin{matrix} \sqrt{1-\gamma} & 0 \\ 0 & \sqrt{1-\delta}\end{matrix}\right) \ \ \ A_1 =  \left( \begin{matrix} 0 & \sqrt{\delta} \\ \sqrt{\gamma} & 0 \end{matrix}\right).
\end{equation}
It was shown in \cite{WPG06} that any such channel is either degradable or anti-degradable. The 
private capacity of such a channel is thus the optimized single-letter coherent information:
\begin{equation}
C_{p}(\cN_{(\gamma,\delta)}) = \nonumber
\max_{t \in [0,1]} \left[H(t(1{-}\gamma){+}(1{-}t)\delta){-}H(t\gamma{+}(1{-}t)\delta)\right].
\end{equation}

This result includes the dephasing and amplitude damping channels as a special case:  setting $\g=0$ 
gives an amplitude damping channel, whereas setting $\g=\d$ gives the bitflip channel (which is
unitarily equivalent to a dephasing channel)\footnote{Note that in \cite{RK04} it was shown that the optimal key rate achievable for dephasing noise of rate $p$ on a maximally correlated classical string
is $1-H(p)$, but because they do not consider general signal states, this does not quite show 
that the private classical capacity of the dephasing channel is the same value, though this formula is implied by our result.}.

The erasure channel with erasure probability $p$, which 
maps $\CC^{d}$ into $\CC^d \oplus \ket{e}$, acts as
\begin{equation}\nonumber
\cN^{\rm erasure}_{(p,d)}(\rho) = (1-p)\rho + p\proj{e}.
\end{equation}
This channel is also degradable, and as a result its private classical capacity is exactly equal to its 
quantum capacity:
\begin{equation}\nonumber
C_p(\cN^{\rm erasure}_{(p,d)}) = (1-2p)\log d.
\end{equation}

\subsection{Independent Phase and Amplitude errors}

The Pauli channel with independent amplitude and phase noise is an
interesting case because of its relation to BB84, and also because it's easy to write as a convex combination of degradables---it's just an equal mixture of two amplitude damping-type channels.

Written explicitly, the channel we are considering is
\begin{eqnarray}\nonumber
\cN_{(q(1{-}q), q^2, q(1{-}q))}(\rho) &=& (1{-}q(2{-}q))\rho + q(1{-}q)X\rho X \nonumber\\
& &  + q^2 Y\rho Y +q(1{-}q)Z\rho Z,\nonumber
\end{eqnarray}
which can also be written as
\begin{equation}
\frac{1}{2}U\cN_{\gamma_q}^{\rm amp damp}(U^\dg \rho U)U^\dg{+}\frac{1}{2}UX\cN^{\rm amp damp}_{\gamma_q}(XU^\dg \rho  UX)XU^\dg,\nonumber
\end{equation}
where $U = e^{i\frac{\pi}{4}X}$ and $\gamma_q = 4q(1-q)$.

From the previous subsection, the private capacity of 
an amplitude damping channel with noise parameter $\gamma$ is
\begin{equation}\nonumber
f(\gamma) = \max_{t \in [0,1]}\left( H(t (1 - \gamma)) - H(t \gamma)\right),
\end{equation}
so that

\begin{equation}\nonumber
C_{p}(\cN_{(q(1-q),q^2,q(1-q) )})  \leq f(\gamma_q).
\end{equation}

\begin{figure}[htbp]
  \centering
  \hspace{1.5cm}
  \includegraphics[width=9cm]{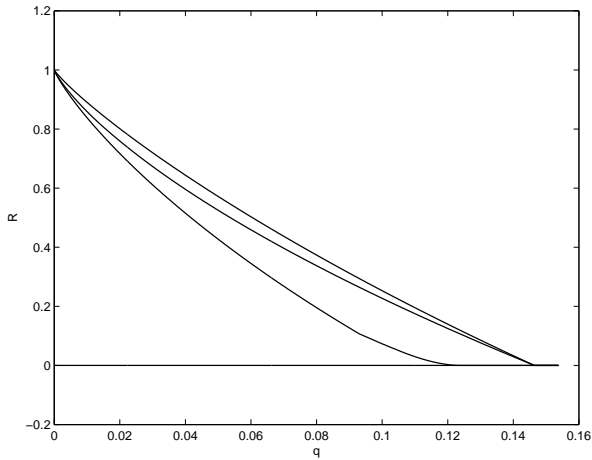}
  \hspace{1.5cm}
  \caption{Bounds on the key rate of BB84 with one-way post-processing as a function of quantum bit error rate, $q$.  The lower curve is the best known achievable key rate from \cite{KGR05,SRS06}.  The upper curve is the ``optimal eavesdropping''' bound on BB84 ({\em without} noisy processing) found 
  in \cite{FGGNP97}, representing the best possible individual attack.  The middle curve is our upper bound from Eq.~(\ref{Eq:BB84UB}).}
  \label{Fig:BB84}
\end{figure}

This gives a threshold  of $\frac{1}{2}\left( 1-\frac{1}{\sqrt{2}}\right)$ beyond which no key can
 be generated, which is the same as found for BB84 in\cite{FGGNP97} , and also confirmed in 
\cite{KGR05} and \cite{MCL06}.  We can also write the $\cN_{(q(1-q),q^2,q(1-q))}$ as a convex combination of dephasing channels with 
dephasing probability $q(2-q)$, which results in slightly tighter bounds for very small noise
(i.e., $q < 0.02$).  Our combined upper bound on key rate is given by
\begin{equation}\label{Eq:BB84UB}
C_{p}(\cN_{(q(1-q),q^2,q(1-q) )})  \leq {\rm conv} \left( 1 - H(q(2-q)), f(\gamma_q)\right),
\end{equation}
and tightens the previous best bounds of \cite{FGGNP97} (which considered only protocols without
noisy processing), and the (straight line)  bound found in \cite{MCL06}
 for all $0 < q <  \frac{1}{2}\left( 1-\frac{1}{\sqrt{2}}\right)$ (see Figure \ref{Fig:BB84}).  The fact 
 that we surpass the bound of \cite{FGGNP97} is particularly interesting, since it is also an
 {\em achievable} key rate against an adversary restricted to individual attacks.  Our bound thus shows 
 that a completely general attack is strictly stronger than an individual attack.

\subsection{Depolarizing channel}

A depolarizing channel with error probability $p$ is a convex combination of six amplitude damping channels with error parameter
\begin{equation}\nonumber
\eta_p = 4\sqrt{1-p}\left( 1 - \sqrt{1-p} \right)
\end{equation}
(see \cite{SSW06} for details).  It is also a convex combination of three dephasing channels  with 
error probability  $p$.  Finally, the secret key capacity is zero whenever a channel is antidegradable, which happens at $p =1/4$ \cite{BDEFMS98}, so that the convexity of $C_{p,ss}$ then implies
\begin{equation}\nonumber
C_{p,ss}(\cN_p) \leq {\rm conv} \left( 1-H(p), f(\eta_p), (1-4p)_+ \right),
\end{equation}
where we have let $x_+ = x$ if $x>0$ and $0$ otherwise.  This expression is equal to the upper
 bound on the depolarizing channel's {\em quantum} capacity
found in \cite{SSW06}, so that the best known upper bounds for this channel actually coincide.

It is worth mentioning that the bound on the threshold for the six-state protocol reported 
in \cite{KGR05} is strictly stronger than the $p=1/4$ threshold implied by our bound.  However, 
the \cite{KGR05} bound {\em does not} apply to the private capacity of the depolarizing channel, since
it is valid only for a restricted set of input states.  

For comparison with the QKD literature, note that the relationship between 
quantum bit error rate, $q$, and depolarizing error probability, $p$,   is $q = 2p/3$.

\subsection{Pauli Channel}
For a general Pauli channel we find the following bound.

\begin{theorem}
Let 
\begin{equation}\nonumber
\cN_{\mathbf p}(\rho) = (1-|{\mathbf p}|)\rho + p_1X\rho X + p_2 Y\rho Y + p_3 Z\rho Z.
\end{equation}
  Then the private classical capacity of $\cN_{\mathbf p}$ satisfies
\begin{equation}\nonumber
C_p (\cN_{\mathbf p}) \leq 1 - H(|{\mathbf p}|),
\end{equation}
where $|{\mathbf p}| = p_1 + p_2 + p_3$.
\end{theorem}
\begin{proof}
Letting $\a_i = p_i/|{\mathbf p}|$, we have
\begin{equation}\nonumber
\cN_{\mathbf p}(\rho) = \a_1\cN^{X}_{|\mathbf p|}(\rho)
+ \a_2\cN^Y_{|\mathbf p|}(\rho)
+ \a_3\cN^Z_{|\mathbf p|},
\end{equation}
where we have let $\cN^X_p(\rho) = (1-p)\rho + pX\rho X $, and similarly for $\cN^Y_p$ and $\cN^Z_p$.  
This is a convex combination of dephasing-like channels with error probability $|\mathbf p|$, which are 
degradable and have a private capacity of $1 - H(|\mathbf p|)$, so that by Theorem \ref{Thm:DMUB} we have the result.
\end{proof}

It is not entirely clear how to best decompose a Pauli channel into a convex combination 
of amplitude damping channels, but it seems likely that such a decomposition 
(or perhaps a decomposition into channels with two Kraus operators) would outperform
the current bound for high noise levels.

\subsection{Relationship to collective attacks in QKD}

In this subsection we describe how the above upper bounds on the private capacity correspond to 
collective attacks on quantum key distribution protocols.  Consider the decomposition of a channel 
$\cN$ into a convex combination of degradable channels, $\cN_i$:
\begin{equation}\nonumber
\cN(\rho) = \sum_i p_i \cN_i,
\end{equation}
where we will call the isometric extension of $\cN_i$ $U_{\cN_i}: A \rightarrow BE$.
The attack associated with this decomposition is as follows:  For each signal state sent, Eve applies $U_{\cN_i}$ with probability $p_i$, sends the $B$ system to Bob, and stores her various $E$ systems until the end of the protocol.  After the protocol is complete, Eve collects all of her $E$ systems associated with $\cN_i$ and (jointly) measures which of the typical eigenvectors of $\rho_{E_i}^{\ox p_i n}$ the state is in.  Because $\cN_i$ is degradable, 
we can calculate exactly how much secret key Alice and Bob can distill from the resulting state---they can get a key rate of exactly $Q^{(1)}(\cN_i)$.  Because a fraction $ p_i$ of the signal states are 
subjected to $\cN_i$, the overall key rate is no more than $\sum_i p_i Q^{(1)}(\cN_i)$.

\section{discussion}
We have studied the capacity of a quantum channel for private classical communication when 
assisted by symmetric channel of an arbitrary size.  For a general channel, we found a single letter formula that, unfortunately, involves an optimization over an auxiliary space that is a prior unbounded.
For degradable channels, we further showed that this optimization can be performed explicitly, and in 
fact the ss-private capacity of such a channel is exactly equal to its single-letter optimized coherent information.  Using this fact, together with the convexity of the ss-capacity for general channels, we
showed how to find upper bounds on the (unassisted) private capacity of a general channel, and provided such bounds for two channels of interest for quantum key distribution.

The most important question we have left unanswered is whether it is possible to bound the dimension
of the symmetric channel necessary to achieve the optimum of the ss-capacity formula found in
Theorem \ref{Thm:SSCAP}.  This could allow very tight bounds on the unassisted capacity. 
In fact, we are unaware of any channel for which the ss-private capacity and unassisted 
private capacity differ, and the conjecture that they are the same is equivalent to the additivity of 
the unassisted capacity, $C_p$.  

We note that for both the independent amplitude and phase noise and the depolarizing channel, the upper bounds are the convex hull of a bound based on decomposition into
dephasing channels, which is strongest in the low noise regime, and a decompostion into amplitude damping channels, which is stronger in the high noise regime.  This suggests that the best collective attacks on quantum key distribution protocols will be qualitatively different in the high and low noise regimes.

It is an interesting question whether there are zero capacity degradable channels 
that are not antidegradable.   This possibility is intriguing, since 
the best known bounds on the zeros of the capacity of most channels come from a no-cloning 
argument (which is essentially to observe that the channel is antidegradable), but these bounds 
are usually not particularly close to the corresponding lower bounds.  Such a channel would also 
be useful for improving estimates on $C_{DM}$ for general channels.

Finally, this work demonstrates (along with \cite{SSW06} ) that assistance from a 
symmetric side channel is a ``nice'' resource, in the sense that it provides a marked 
simplification over the unassisted case for the private capacity.  Further examples of nice 
resources are free EPR pairs,  which lead to the single-letter formula for the entanglement 
assisted capacity of \cite{BSST02,AC97}, 
and PPT-preserving
operations, which dramatically simplify the theory of entanglement manipulations \cite{EVWW01,Rains-PPT}.  What are the other ``nice'' resources?

\section*{Acknowledgments}
I am grateful to  Debbie Leung, John Smolin, Andreas Winter,  and Charlie Bennett for helpful conversations, to the Institute for Quantum Computing at the University of Waterloo, where this work was initiated, and the  United Kingdom  Engineering and Physical Sciences Research Council for financial support.

\bibliographystyle{apsrev}

\end{document}